\documentclass[aps,preprint,pra]{revtex4}
\usepackage{graphicx}
\usepackage[pdftex]{color}

\begin{document}

\title{A reasonable thing that just might work}\thanks{To appear in {\it Quantum Nonlocality and Reality: 50 Years of Bell`s theorem}, eds. S. Gao and M. Bell (Cambridge U. Press), 2015, in press}

\author{Daniel Rohrlich}
\affiliation{Department of Physics, Ben Gurion University of the Negev, Beersheba
8410501 Israel}

\date{\today}

\begin{abstract}

In 1964, John Bell proved that quantum mechanics is ``unreasonable" (to use Einstein's term):  there are nonlocal bipartite quantum correlations.  But they are not the most nonlocal bipartite correlations consistent with relativistic causality (``no superluminal signalling"):  also maximally nonlocal ``superquantum" (or ``PR-box") correlations are consistent with relativistic causality.  I show that---unlike quantum correlations---these correlations do not have a classical limit consistent with relativistic causality.  The generalization of this result to all stronger-than-quantum nonlocal correlations is a derivation of Tsirelson's bound---a theorem of quantum mechanics---from the three axioms of relativistic causality, nonlocality, and the existence of a classical limit.  But is it reasonable to derive (a part of) quantum mechanics from the unreasonable axiom of nonlocality?!  I consider replacing the nonlocality axiom with an equivalent axiom that even Bell and Einstein might have considered reasonable:  an axiom of local retrocausality.

\end{abstract}

\maketitle

In 1964, John Bell \cite{bell} proved that quantum mechanics is ``unreasonable", as defined by Einstein, Podolsky and Rosen \cite{EPRB} in 1935: ``No reasonable definition of reality could be expected to permit this."  ``This" (i.e. violation of ``Einstein separability" to use a technical term, or ``spooky action at a distance" as Einstein put it) turns out to be endemic to quantum mechanics. For example, pairs of photons measured at spacelike separations may yield nonlocal quantum correlations, i.e. correlations that cannot be traced to any data or ``programs" the photons carry with them.  As Bell \cite{jb} put it 20 years later, ``For me, it is so reasonable to assume that the photons in those experiments carry with them programs, which have been correlated in advance, telling them how to behave.  This is so rational that I think that when Einstein saw that, and the others refused to see it, {\it he} was the rational man.  The other people, although history has justified them, were burying their heads in the sand.  I feel that Einstein's intellectual superiority over Bohr, in this instance, was enormous; a vast gulf between the man who saw clearly what was needed, and the obscurantist.  So for me, it is a pity that Einstein's idea doesn't work.  The reasonable thing just doesn't work." True, history has confirmed nonlocal quantum correlations; but has history passed judgment on Einstein and ``the others"?  Consider what Newton \cite{newton} wrote about his own theory of gravity:  ``That gravity should be innate, inherent and essential to matter so that one body may act upon another at a distance through a vacuum without the mediation of anything else, by and through which their action or force may be conveyed from one to another, is to me so great an absurdity that I believe no man who has in philosophical matters any competent faculty of thinking can ever fall into it."  More than two centuries years later, Einstein confirmed Newton's misgivings:  gravitational interactions are indeed local.  Einstein's theory of gravity is free of the ``absurdity" of action at a distance.  Well, if it took centuries for history to justify Newton's rejection of action at a distance, couldn't history yet justify Einstein's rejection of action at a distance?

This paper offers, not {\it the} reasonable thing, but $a$ reasonable thing that just might work.  Section I reviews the search for simple physical axioms from which to derive quantum mechanics.  Ideally, such a search could help us understand the theory and make it seem more reasonable.  But, while we can derive a part of quantum mechanics from three simple physical axioms, one of the three axioms is the unreasonable axiom of nonlocality!  Apparently, we are no better off than before.  However, Sect. II considers replacing the axiom of nonlocality with an axiom that even Bell and Einstein might have considered reasonable:  an axiom of local retrocausality (microscopic time-reversal symmetry).  Then Sect. III rewrites the derivation in Sect. I using the three axioms of causality, local retrocausality, and the existence of a classical limit.
\section{Nonlocal correlations in the classical limit}
\label{section1}
It is convenient to discuss Bell's inequality in the form derived by Clausner, Horne, Shimony and Holt \cite{CHSH} (CHSH) for spacelike separated measurements by  ``Alice" and ``Bob" on a bipartite system:
\begin{equation}
\left\vert C(a,b) +C(a,b^\prime) +C(a^\prime,b) - C(a^\prime,b^\prime) \right\vert\le 2~~~,
\label{BCHSH}
\end{equation}
where $a, ~a^\prime, ~b$ and $b^\prime$ are observables with eigenvalues in the range $[-1,1]$; Alice measures $a$ or $a^\prime$ in her laboratory, Bob measures $b$ or $b^\prime$ in his laboratory, and the correlations $C(a,b)$, etc., emerge from their measurements.  Correlations that violate Eq.\ (\ref{BCHSH}) by {\it any} amount are nonlocal.  But it is a curious fact, discovered by Tsirelson \cite{ts}, that the violation of Eq.\ (\ref{BCHSH}) by quantum correlations $C_Q(a,b)$, etc., is bounded by 2$\sqrt{2}$:
\begin{equation}
\left\vert C_Q(a,b) +C_Q(a,b^\prime) +C_Q(a^\prime,b) - C_Q(a^\prime,b^\prime) \right\vert\le 2\sqrt{2}~~~,
\label{TB}
\end{equation}
even though it is straightforward to define ``superquantum" correlations
\begin{equation}
C_{SQ}(a,b) = C_{SQ}(a,b^\prime) = C_{SQ}(a^\prime,b) = 1 = -C_{SQ}(a^\prime,b^\prime)
\label{SQ}
\end{equation}
that violate Eq.\ (\ref{BCHSH}) maximally.  A good guess is that superquantum (or ``PR-box" \cite{PR}) correlations are too strong to be consistent with relativistic causality, but this guess is easily disproved \cite{PR,and}.  Just assume that when Alice measures $a$ or $a^\prime$ she gets $\pm 1$ with equal probability, and likewise when Bob measures $b$ or $b^\prime$; this assumption is consistent with Eq.\ (\ref{SQ}) and it implies that Alice and Bob cannot signal to each other, since in any case Alice and Bob obtain $\pm 1$ with equal probability from their measurements.

Nonlocal quantum correlations are unreasonable, and not even maximally unreasonable!

But perhaps we should not be so surprised that PR-box correlations are consistent with relativistic causality.  After all, we have set up quite an artificial comparison.  We have not compared two theories.  We have compared nonlocal quantum correlations belonging to a complete theory---quantum mechanics---with ad hoc super-duper nonlocal correlations that do not belong to any theory we know.  We are not even comparing apples and oranges.  We are comparing a serial Nobel prize winner and a lottery winner.  Quantum mechanics, as a complete theory, is subject to constraints.  In particular, quantum mechanics has a classical limit.  In this limit there are no complementary observables; there are only macroscopic observables, all of which are  jointly measurable.  This classical limit---our direct experience---is an inherent constraint, a kind of boundary condition, on quantum mechanics and on any generalization of quantum mechanics.  Thus stronger-than-quantum correlations, too, must have---as a minimal requirement---a classical limit.

And now the fun begins \cite{Y80}.  Consider the PR box and note that if Alice measures $a$ and obtains 1, she can predict with certainty that Bob will obtain 1 whether he measures $b$ or $b^\prime$; if she obtains $-1$,  she can predict with certainty that he will obtain $-1$ whether he measures $b$ or $b^\prime$.  (By contrast, quantum correlations would allow Alice to predict with certainty only the result of measuring $b$ or the result of measuring $b^\prime$ but not both \cite{EPRB}.)  If Alice measures $a^\prime$, she can predict with certainty that Bob will obtain her result if he measures $b$ and the opposite result if he measures $b^\prime$.  Thus, all that protects relativistic causality is the (tacitly assumed) complementarity between $b$ and $b^\prime$:  Bob cannot measure both, although---from Alice's point of view---no uncertainty principle governs $b$ and $b^\prime$.

Next, suppose that Alice measures $a$ or $a^\prime$ consistently on $N$ pairs.  Let us define macroscopic observables $B$ and $B^\prime$:
\begin{equation}
B={{b_1+ b_2 +\dots+b_N}\over N}~~~~,~~~B^\prime={{b^\prime_1+ b^\prime_2 +\dots+
b^\prime_N}\over N}~~~,
\label{defbbp}
\end{equation}
where $b_m$ and $b^\prime_m$ represent $b$ and $b^\prime$, respectively, on the $m$-th pair. Alice already knows the values of $B$ and $B^\prime$, and there must be $some$ measurements that Bob can make to obtain partial information about $both$ $B$ and $B^\prime$; for, in the classical limit, there can be no complementarity between $B$ and $B^\prime$.  Now it is true that $a=1$ and $a=-1$ are equally likely, and so the average values of $B$ and $B^\prime$ vanish, whether Alice measures $a$ or $a^\prime$.  But if she measures $a$ on each pair, then typical values of $B$ and $B^\prime$ will be $\pm 1/\sqrt{N}$ (but possibly as large as $\pm 1$) and correlated.  If she measures $a^\prime$ on each pair, then typical values of $B$ and $B^\prime$ will be $\pm 1/\sqrt{N}$ (but possibly as large as $\pm 1$) and $anti$-correlated.  Thus Alice can signal a single bit to Bob by consistently choosing whether to measure $a$ or $a^\prime$.  This claim is delicate because the large-$N$ limit in which $B$ and $B^\prime$ commute is also the limit that suppresses the fluctuations of $B$ and $B^\prime$.  We cannot make any assumption about the $approach$ to the classical limit; all that we assume is that it exists, e.g. that the uncertainty product $\Delta B \Delta B^\prime$ can be made as small as desired, for large enough $N$.  On the other hand, the axiom of relativistic causality cannot grant Bob even the slightest indication about both $B$ and $B^\prime$.  Hence all we need is that when Bob detects a correlation, it is more likely that Alice measured $a$ than when he detects an anti-correlation.  If it were not more likely, it would mean that Bob's measurements yield zero information about $B$ or about $B^\prime$, contradicting the fact that there is a classical limit in which $B$ and $B^\prime$ are jointly measurable.

To ensure that Bob has a good chance of measuring $B$ and $B^\prime$ accurately enough to determine whether they are correlated or anti-correlated, $N$ may have to be large and therefore the fluctuations in $B$ and $B^\prime$ will be small.  However, Alice and Bob can repeat this experiment (on $N$ pairs at a time) as many times as it takes to give Bob a good chance of catching and measuring large enough fluctuations.  Alice and Bob's expenses and exertions are not our concern. Relativistic causality does not forbid superluminal signalling only when it is cheap and reliable.  Relativistic causality forbids superluminal signalling altogether.

For example, let us suppose Bob considers only those sets of $N$ pairs in which $B=\pm1$ and $B^\prime=\pm 1$.  The probability of $B=1$ is $2^{-N}$.  But if Alice is measuring $a$ consistently, the probability of $B=1$ $and$ $B^\prime =1$ is also $2^{-N}$, and not $2^{-2N}$, while the probability of $B=1$ and $B^\prime =-1$ vanishes.  If Alice is measuring $a^\prime$ consistently, the probabilities are reversed.  (These probabilities must be folded with the scatter in Bob's measurements, but the scatter is independent of what Alice measures.)  Thus with unlimited resources, Alice can send a (superluminal) signal to Bob. Superquantum (PR-box) correlations are {\it not} consistent with relativistic causality in the classical limit.

We have ruled out superquantum correlations \cite{Y80}.  To derive quantum correlations, however, we have to rule out all stronger-than-quantum correlations, i.e. we have to derive Tsirelson's bound from the three axioms of nonlocality, relativistic causality, and the existence of a classical limit.  The proof appears elsewhere \cite{max}.

The existence of a classical limit is not the only axiom we can consider adding to
the axioms of nonlocality and relativistic causality.  Alternative axioms \cite{o} (or a stronger axiom of relativistic causality called ``information causality" \cite{ic}), have been shown to rule out PR-box correlations, and come close to ruling out all stronger-than-quantum correlations.  However, the physical significance of these axioms requires clarification.  Navascu{\'e}s and Wunderlich \cite{nw} consider an axiom for a classical limit, but define the classical limit via the ``wiring" \cite{bs} of entangled systems, and not via complementary measurements that become jointly measurable as the number $N$ of systems grows without bound.
\section{Local retrocausality as an axiom}
\label{section2}
Bell's theorem rules out any locally causal account of quantum mechanics.  But a number of authors, most notably Price \cite{price}, have suggested a locally causal-retrocausal account of quantum mechanics.  Here ``causality" means ``relativistic causality" as before (i.e. no superluminal signalling); what is ``retro" is that the effect precedes the cause.  If the retrocausality is {\it local}---no action at a distance---then the order of cause and effect is independent of the reference frame.  Retrocausality is an expression of a fundamental time-reversal symmetry in physics.  While time-reversal symmetry is not manifest at the macroscopic level---for example, a star emits more light than it absorbs---we explain the asymmetry by saying that the universe has not reached a state of maximum entropy.  At the same time, almost all fundamental physical processes at the microscopic level exhibit time-reversal symmetry.  Aharonov, Bergmann and Lebowitz (ABL) \cite{abl} derived an explicitly time-symmetric formula for intermediate quantum probabilities, conditioned on initial and final states; they suggested that quantum mechanics has no arrow of time of its own and that time asymmetry (e.g. in measurements) originates in macroscopic physics.  While the ABL formula is not manifestly local, it opens the way to a local account of quantum mechanics via local retrocausality.  Such an account would replace nonlocality with something not only local, but even palatable:  a fundamental time-reversal symmetry of microscopic causality and retrocausality.  Moreover, if the account includes the quantum correlations that violate Bell's inequality, we can replace the axiom of nonlocality assumed in Sect. I with an axiom of local retrocausality, and try to derive quantum mechanics from the three axioms of (relativistic) causality, local retrocausality, and the existence of a classical limit.  Sect. III begins such a derivation.

\begin{figure}
\centerline{
\includegraphics*[width=120mm]{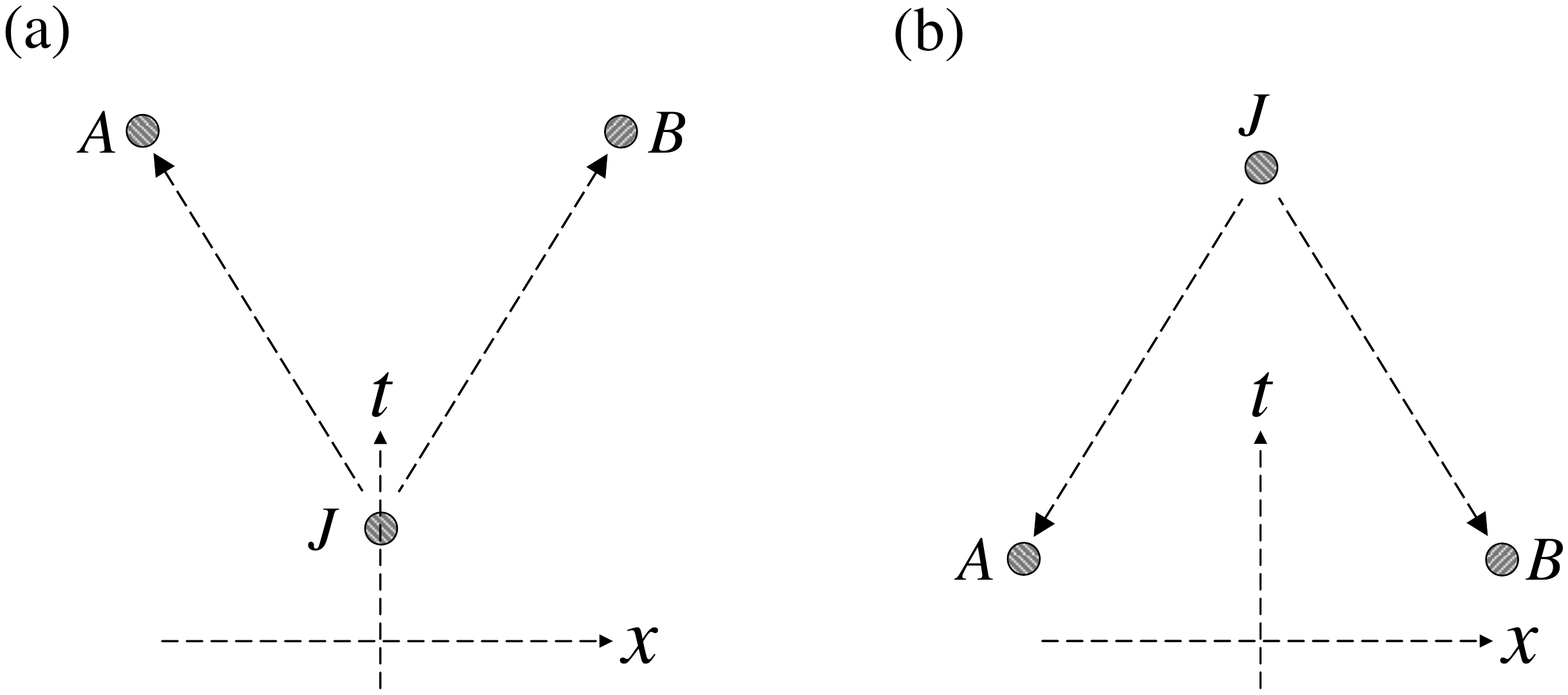}}
\caption[]{Configurations in which Jim can (a) causally and (b) retrocausally put pairs of particles shared by Alice and Bob in product or entangled states, as he chooses.  The dashed arrows connect cause with effect.}
\label{Fig1ab}
\end{figure}
Remarkably, retrocausality is intrinsic to quantum mechanics, as we see if we consider three observers, Alice, Bob and Jim, who share an ensemble of triplets of spin-1/2 particles in the Greenberger, Horne and Zeilinger (GHZ) \cite{GHZ} state $\vert \rm{GHZ}\rangle = (\vert \uparrow_A \uparrow_B \uparrow_J\rangle - \vert \downarrow_A \downarrow_B \downarrow_J\rangle )/\sqrt{2}$.  (See Fig. 1; Alice, Bob and Jim each get one particle in each triplet.)  Let Alice and Bob, at spacetime points $A$ and $B$, measure spin components ${\bf \sigma}^{(A)} \cdot {\bf {\hat n}}_A$ and ${\bf \sigma}^{(B)} \cdot {\bf {\hat n}}_B$, respectively, on their particles.  For simplicity, let the unit vectors ${\bf {\hat n}}_A$ and ${\bf {\hat n}}_B$ (which may change from particle to particle) lie in the $xy$-plane.  Let Jim have the special role of the ``jammer" \cite{jam}; he chooses whether to put the particles held by Alice and Bob in a product state or an entangled state.  To put them in a product state, Jim (at spacetime point $J$) measures $\sigma_z^{(J)}$, the $z$-component of the spin of his particles.  To put them in an entangled state, he measures, say, $\sigma_x^{(J)}$.  It doesn't matter when Jim makes his measurements.  In Fig. 1(a), Jim's measurements precede Alice's and Bob's by a timelike separation; but in Fig. 1(b), Alice's and Bob's measurements precede Jim's by a timelike separation.  Either way, Jim cannot send a superluminal signal to Alice and Bob, because his measurements leave the pairs held by Alice and Bob in a mixed state---either a mixture of the product states $\vert \uparrow_A \uparrow_B\rangle$ and $\vert \downarrow_A \downarrow_B\rangle$ or a mixture of the entangled states $(\vert \uparrow_A \uparrow_B \rangle - \vert \downarrow_A \downarrow_B \rangle )/\sqrt{2}$ and $(\vert \uparrow_A \uparrow_B \rangle + \vert \downarrow_A \downarrow_B \rangle )/\sqrt{2}$.  Without access to the {\it results} of Jim's measurements, Alice and Bob cannot distinguish between these two mixtures.  But {\it with} access to the results, they can bin their data accordingly and verify that their results either do or do not violate Bell's inequality in accordance with whether Jim chooses to entangle their pairs or not.

Thus, Fig. 1(b) nicely illustrates the fact that quantum mechanics is retrocausal---even if we take quantum nonlocality at face value without considering retrocauality.  On the one hand, there is no reason to doubt that Alice, Bob, and Jim have free will.  Indeed, the results of Alice and Bob's measurements are consistent with whatever Jim chooses, right up to the moment when he decides to measure $\sigma_z^{(J)}$ or $\sigma_x^{(J)}$ on each of his particles and record the results.  On the other hand, there is no doubt about the effect (in Jim's past light cone) of Jim's choice.  After Alice and Bob obtain the results of Jim's measurements (within his forward light cone) they can reconstruct from their data whether their particles were entangled or not at the time they measured them.  Thus quantum mechanics is retrocausal (though not necessarily {\it locally} retrocausal).

\begin{figure}
\centerline{
\includegraphics*[width=120mm]{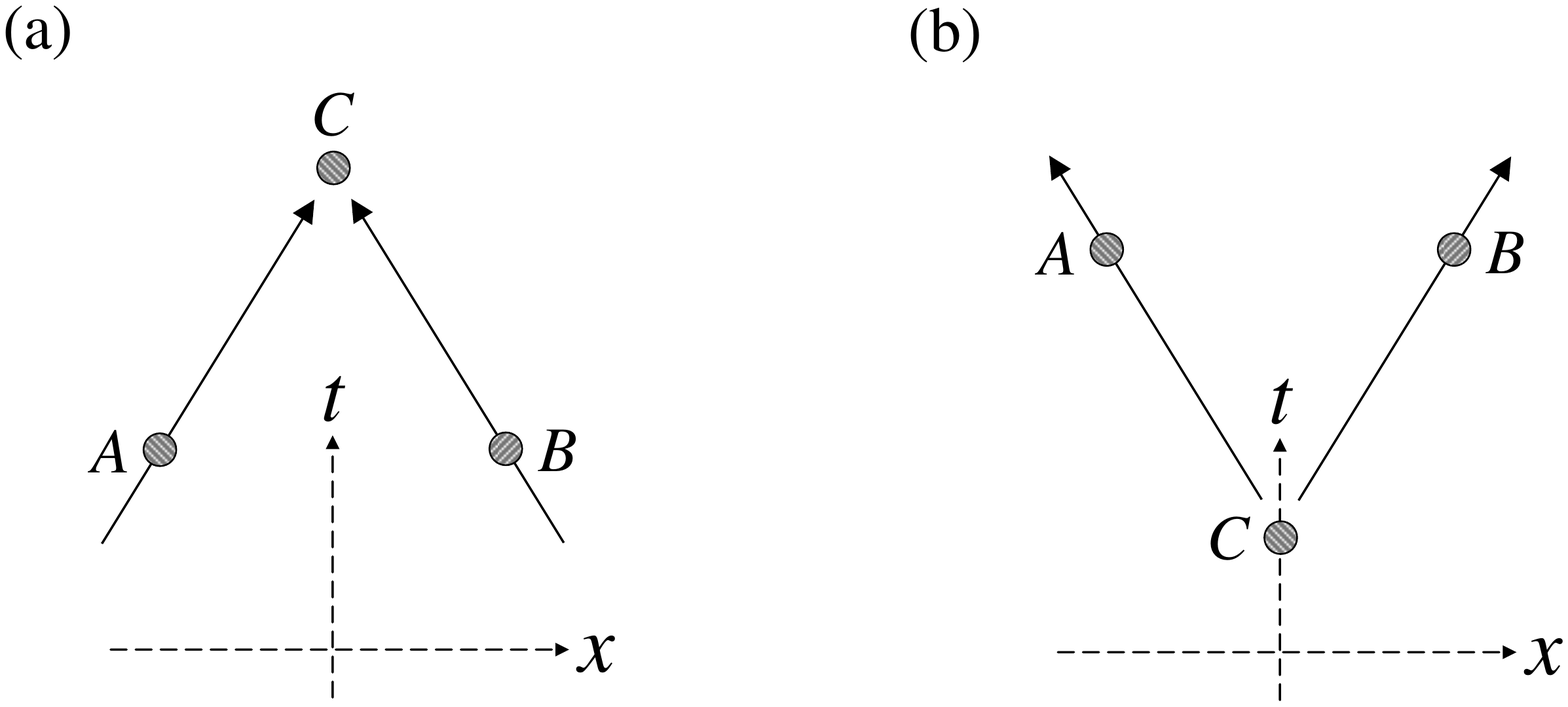}}
\caption[]{(a) Alice and Bob independently measure (prepare) spin states of a pair of converging spin-1/2 particles; when the particles meet, Claire's measurement leaves them in a product state or a Bell state. (b) In the reverse sequence, Claire's measurement prepares a product state or a Bell state of the two (diverging) particles, for Alice and Bob to independently measure.}
\label{Fig2ab}
\end{figure}
Now, to illustrate how local retrocausality could obviate nonlocality, consider Fig. 2.  Figure 2(a) represents what might be called a ``reverse EPR experiment".  Alice and Bob, at spacetime points $A$ and $B$, measure spin components ${\bf \sigma}^{(A)} \cdot {\bf {\hat n}}_A$ and ${\bf \sigma}^{(B)} \cdot {\bf {\hat n}}_B$, respectively, on spin-1/2 particles approaching in pairs, in opposite directions, along the $x$ axis. For each pair, Alice and Bob are completely free to choose the unit vectors ${\bf \hat n}_A$ and ${\bf \hat n}_B$ independently.  The spin states of the particles before they reach Alice and Bob are irrelevant, and the pairs leave in products of eigenstates of ${\bf \sigma}^{(A)} \cdot {\bf {\hat n}}_A$ and ${\bf \sigma}^{(B)} \cdot {\bf {\hat n}}_B$ with eigenvalues $\pm 1$.  Let  $\vert\xi_A \xi^\prime_B \rangle$ denote these product states.  At the point where the particles meet, Claire is free to make one of two measurements.  She either measures a nondegenerate operator $P$ with the product eigenstates $\vert\uparrow_A\uparrow_B \rangle$, $\vert\downarrow_A\uparrow_B \rangle$, $\vert\uparrow_A\downarrow_B \rangle$, and $\vert\downarrow_A\downarrow_B \rangle$ (i.e. products of the eigenstates of $\sigma^{(A)}_z$ and $\sigma^{(B)}_z$), or she measures a nondegenerate operator $B$ with the Bell states $\left( \vert \uparrow_A \uparrow_B)\rangle \pm \vert \downarrow_A \downarrow_B \right) / {\sqrt{2}}$ and $\left( \vert \uparrow_A \downarrow_B \rangle \pm \vert \uparrow_A \downarrow_B \right) / {\sqrt{2}}$ as eigenstates.  The Bell states are entangled.  Let Alice and Bob send Claire the results of their measurements.  Suppose Claire chooses to measure the Bell operator each time; she can bin the data for each pair that arrives in her laboratory according to the Bell state that she finds for it.  Over time, she will be able to measure the quantum correlations between Alice's and Bob's measurements from the binned data, for each Bell state.  These quantum correlations are precisely the nonlocal quantum correlations that violate Bell's inequality, since for quantum probabilities, and hence for correlations, time order does not matter:  $\vert \langle \xi_A \xi^\prime_B \vert B_i\rangle \vert^2$ = $\vert \langle B_i \vert \xi_A \xi^\prime_B \rangle \vert^2$, where $\vert B_i \rangle$ is any one of the four Bell states.  Yet nothing even slightly nonlocal is going on here. The results of Alice's and Bob's measurements propagate locally and causally to Claire, who ``clarifies" the overall state of each pair of particles that arrives in her laboratory with her measurement.  The reason that Fig. 2(a) is locally causal is that local causality brings the results of Alice's, Bob's and Claire's measurements all together at the spacetime point $C$.  In Fig. 2(b), local causality {\it cannot} bring the results of Alice's, Bob's and Claire's measurements all together at any point, because the particles in each pair diverge to spacetime points $A$ and $B$.  Thus the conditions for Bell's theorem hold and the quantum correlations of Fig. 2(b), which are also the quantum correlations of Fig. 2(a), are nonlocal.

Nevertheless, time-reversal symmetry suggests that Fig. 2(a) and Fig. 2(b) are analogous.  Perhaps local retrocausality could play the role in Fig. 2(b) that local causality cannot play:  local retrocausality could propagate the results of Alice's and Bob's measurements at $A$ and $B$, respectively, {\it backwards} in time to bring the results of Alice's, Bob's and Claire's measurements all together at the spacetime point $C$.  Using the ABL formula, we can express the conditional probability that Claire's measurement at spacetime point $C$ yields the Bell state $\vert B_j\rangle$ as
\begin{equation}
{\rm prob} (\vert B_j\rangle) = {{\vert\langle \xi_A\xi^\prime_B \vert U(t_{AB}-t_C)  \vert B_j\rangle \langle B_j\vert U(t_C- t_0)\vert 0\rangle\vert^2}
\over {\sum_{i=1}^4
\vert\langle \xi_A\xi^\prime_B \vert U(t_{AB}-t_C) \vert B_i\rangle \langle B_i\vert U(t_C- t_0)\vert 0\rangle\vert^2}}~~~,
\label{abl1}
\end{equation}
where $\vert 0\rangle$ is the state of the two spin-1/2 particles at time $t_0$, before Claire's measurement, $t_C$ is the time of her measurement, and $t_{AB}$ is the time of Alice's and Bob's measurements, which we can take (for simplicity and without lost of generality) to be simultaneous.  The unitary operator $U(t_{AB}-t_C)$ can be rewritten as $U^\dagger (t_C-t_{AB})$ to remove any arrow of time from the ABL formula:
\begin{equation}
{\rm prob} (\vert B_j\rangle) = {{\vert\langle \xi_A\xi^\prime_B \vert U^\dagger (t_C-t_{AB}) \vert B_j\rangle \langle B_j\vert U(t_C- t_0)\vert 0\rangle\vert^2}
\over {\sum_{i=1}^4
\vert\langle \xi_A\xi^\prime_B \vert U^\dagger (t_C-t_{AB})) \vert B_i\rangle \langle B_i\vert U(t_C- t_0)\vert 0\rangle\vert^2}}~~~;
\label{abl2}
\end{equation}
here the initial state $\vert 0\rangle$ and final state $\vert \xi_A \xi^\prime_B \rangle$ both evolve {\it locally} towards the intermediate time of Claire's measurement.

The ABL formula realizes the time-reversal symmetry between Fig. 2(a) and Fig. 2(b).  But we have already noted that time-reversal symmetry holds only for microscopic physics, and not for macroscopic physics.  In particular, the Born rule belongs to the realm of macroscopic physics.  In Fig. 2(b), we can use $\vert \langle \xi_A \xi^\prime_B \vert B_i\rangle \vert^2$ = $\vert \langle B_i \vert \xi_A \xi^\prime_B \rangle \vert^2$ to predict the probability of the state $\vert  \xi_A \xi^\prime_B \rangle$ given the state $\vert B_i\rangle$, we {\it cannot} use it to retrodict the probability of the state $\vert B_i\rangle$ given the state $\vert  \xi_A\xi^\prime_B \rangle$.  More concretely, Claire in Fig. 2(b) could certainly entangle two spin-1/2 particles by measuring on them an operator $B$ with the Bell states as eigenstates; but in another experiment to test Bell's inequality, the particles might be photons in a singlet state, produced by the decay of an excited state of an atom.  If so, the time-reversed experiment---Fig. 2(a) in which the photons converge so precisely as to excite an atom---is much less likely.  However, the ABL approach \cite{abl,ar} is still valid:  quantum mechanics (microscopic physics) contains no arrow of time, and the macroscopic arrow of time derives from thermodynamics and boundary conditions on the universe.  If so, perhaps we can overlook the imperfect analogy between Fig. 2(a) and Fig. 2(b) and let retrocausality evolve the states at $A$ and $B$ back to $C$, where quantum probabilities determine the actual sequence of results.  This retrocausal description fits naturally with the ``two-state-vector" formulation of quantum mechanics \cite{abl,update}.

It is also consistent with free will, in the following sense.  There would be a problem regarding free will if, say, Alice could obtain any information about what she measured {\it before} the measurement.  Any physical theory that allowed such a causal loop would be inconsistent.  But suppose Alice could not obtain any such information before the measurement, but someone else could.  No causal loop could arise, but would we still say that Alice has free will?  The question does not apply to Fig. 2(b) because no one has access to information about Alice's measurement before the event $A$:  a normal (``strong") measurement between $A$ and $C$ would eliminate the causal/retrocausal connection between the two events, and a ``weak" measurement \cite{weak} could yield a result only {\it after} Alice's measurement.
\section{PR-box correlations from local retrocausality}
\label{section3}
We can now define a toy model for the PR box as a retrocausal box rather than a nonlocal box (as Argaman \cite{arg} defined a toy model for bipartite singlet correlations).  Returning to Fig. 2(b), we let Alice's and Bob's choices of what to measure ($a$ or $a^\prime$ and $b$ or $b^\prime$, respectively) propagate retrocausally to $C$, where (for the PR box) choices $(a,b)$, $(a,b^\prime)$ and $(a^\prime, b)$ yield values $(1,1)$ or $(-1,-1)$ with equal probability, while choice $(a^\prime, b^\prime)$ yields values $(1,-1)$ or $(-1,1)$ with equal probability.  (By analogy with the previous section, we could let Claire clarify if the box is a PR box or a different but equivalent box.)  Then Alice and Bob's measured correlations are PR-box correlations, i.e. the retrocausal box is equivalent to the nonlocal box.  And now the conclusions of Sect. I apply to the retrocausal box just as they apply to the nonlocal box:  Alice and Bob can violate the axiom of no superluminal signalling in the classical limit (the limit of arbitrarily many boxes).  In other words, the PR-box is {\it not} causal in the classical limit.  Just as in Sect. I, we can eliminate PR-box correlations as not satisfying the three axioms of causality, local retrocausality and the existence of a classical limit.  Likewise, from these three axioms alone we can expect to derive Tsirelson's bound---a theorem of quantum mechanics.

To conclude, local retrocausality offers us an alternative to ``spooky action at a distance".  Would Einstein have accepted it?  Is local retrocausality a deep principle worthy of being an axiom?  It is appropriate to let Bell have the last word \cite{jb2}:

``I think Einstein thought that Bohm's model was too glib---too simple. I think he was looking for a much more profound rediscovery of quantum phenomena. The idea that you could just add a few variables and the whole thing [quantum mechanics] would remain unchanged apart from the interpretation, which was a kind of trivial addition to ordinary quantum mechanics, must have been a disappointment to him. I can understand that---to see that that is all you need to do to make a hidden-variable theory. I am sure that Einstein, and most other people, would have liked to have seen some big principle emerging, like the principle of relativity, or the principle of the conservation of energy.  In Bohm's model one did not see anything like that."

\begin{acknowledgments}
I thank Huw Price and Ken Wharton for stimulating correspondence and Yakir Aharonov, Sandu Popescu and Nathan Argaman for critical comments.  I acknowledge support from the John Templeton Foundation (Project ID 43297) and from the Israel Science Foundation (grant no. 1190/13).  The opinions expressed in this publication are mine and do not necessarily reflect the views of either of these supporting foundations.

\end{acknowledgments}


\begin{references}

\bibitem{bell} J. S. Bell, ``On the Einstein Podolsky Rosen paradox", {\it Physics} {\bf 1}, 195 (1964).

\bibitem{EPRB}A. Einstein, B. Podolsky and N. Rosen, ``Can quantum-mechanical description of reality be considered complete?" {\it Phys. Rev.} {\bf 47}, 777 (1935); see also N. Bohr, ``Can quantum-mechanical description of reality be considered complete?" {\it Phys. Rev.} {\bf 48}, 696 (1935).

\bibitem{jb}{J. Bernstein, {\it Quantum Profiles} (Princeton:  Princeton U. Press), 1991, p. 84.}

\bibitem{newton}{I. Newton, letter to R. Bentley, 25 February 1693, in {\it The Correspondence of Isaac Newton}, Vol. {\bf III}, ed. H. W. Turnbull (Cambridge: Cambridge U. Press), 1961, pp. 253-56, cf. p. 254; punctuation and spelling edited.}

\bibitem{CHSH} J. F. Clauser\index{Clauser, John F.}, M. A. Horne\index{Horne, Michael A.}, A. Shimony\index{Shimony, Abner}, and R. A. Holt\index{Holt, Richard A.}, ``Proposed experiment to test local hidden-variable theories", {\it Phys. Rev. Lett.} {\bf 23}, 880 (1969).

\bibitem{ts}B. S. Tsirelson (Cirel'son), ``Quantum generalizations of Bell's inequality", {\it Lett. Math. Phys.} {\bf 4}, 93 (1980).

\bibitem{PR} S. Popescu\index{Popescu, Sandu} and D. Rohrlich\index{Rohrlich, Daniel}, ``Quantum nonlocality as an axiom", {\it Found. Phys.} {\bf 24}, 379 (1994).

\bibitem{and} L. Khalfin and B. Tsirelson, ``Quantum and quasi-classical analogs of Bell inequalities", in {\it Symposium on the Foundations of Modern Physics '85}, P. Lahti et al., eds. (Singapore:  World-Scientific), 1985, p. 441; P. Rastall, ``Locality, Bell's theorem, and quantum mechanics", {\it Found. Phys.} {\bf 15}, 963 (1985); G. Krenn and K. Svozil, ``Stronger-than-quantum correlations", {\it Found. Phys.} {\bf 28}, 971 (1998).

\bibitem{Y80} D. Rohrlich\index{Rohrlich, Daniel}, ``PR-box correlations have no classical limit", in {\it Quantum Theory: A Two-Time Success Story} [Yakir Aharonov\index{Aharonov, Yakir} Festschrift], eds. D. C. Struppa and J. M. Tollaksen (Milan: Springer), 2013, pp. 205-211.

\bibitem{max} D. Rohrlich\index{Rohrlich, Daniel}, ``Stronger-than-quantum bipartite correlations violate relativistic causality in the classical limit", arXiv:1408.3125; A. Carmi and D. Rohrlich\index{Rohrlich, Daniel}, in preparation.  See also N. Gisin\index{Gisin, Nicolas}, ``Quantum measurement of spins and magnets, and the classical limit of PR-boxes", arXiv:1407.8122.

\bibitem{o} W. van Dam, {\it  Nonlocality \& Communication Complexity} (Ph.D. thesis), Oxford University (2000); ``Implausible consequences of superstrong nonlocality", quant-ph/0501159 (2005); D. Dieks, ``Inequalities that test locality in quantum mechanics", {\it Phys. Rev.} {\bf A66}, 062104 (2002); H. Buhrman and S. Massar, ``Causality and Tsirelson's bounds", {\it Phys. Rev.} {\bf A72}, 052103 (2005);  J. Barrett and S. Pironio, ``Popescu-Rohrlich correlations as a unit of nonlocality", {\it Phys. Rev. Lett.} {\bf 95}, 140401 (2005); G. Brassard, H. Buhrman, N. Linden, A. A. M{\'e}thot, A. Tapp and F. Unger, ``Limit on nonlocality in any world in which communication complexity is not trivial", {\it Phys. Rev. Lett.} {\bf 96}, 250401 (2006); J. Barrett, ``Information processing in generalized probabilistic theories", {\it Phys. Rev.} {\bf A75}, 032304 (2007); D. Gross, M. M{\"u}ller, R. Colbeck and O. C. O. Dahlsten, ``All reversible dynamics in maximally nonlocal theories are trivial", {\it Phys. Rev. Lett.} {\bf 104}, 080402 (2010).

\bibitem{ic}M. Paw{\l}owski et al., ``Information causality as a physical principle", {\it Nature} {\bf 461}, 1101 (2009).

\bibitem{nw}M. Navascu{\'e}s and H. Wunderlich, ``A glance beyond the quantum model", {\it Proc. R. Soc. A} {\bf 466}, 881 (2010).

\bibitem{bs}N. Brunner and P. Skrzypczyk, ``Nonlocality distillation and postquantum theories with trivial communication complexity", {\it Phys. Rev. Lett.} {\bf 102}, 160403 (2009).

\bibitem{price} H. Price\index{Price, Huw}, {\it Time's Arrow \& Archimedes' Point:  New Directions for the Physics of Time} (New York: Oxford U. Press), 1996; P. W. Evans, H. Price\index{Price, Huw} and K. B. Wharton\index{Wharton, Ken}, ``New slant on the EPR-Bell experiment", {\it Brit. J. Phil. Sci.} {\bf 64}, 297 (2013); H. Price\index{Price, Huw} and K. Wharton\index{Wharton, Ken}, ``Dispelling the quantum spooks---a clue that Einstein missed?", arXiv:1307.7744v1.

\bibitem{abl} Y. Aharonov\index{Aharonov, Yakir}, P. G. Bergmann and J. L. Lebowitz, ``Time symmetry in the quantum process of measurement", {\it Phys. Rev.} {\bf 134} (1964) B1410.  See also Y. Aharonov\index{Aharonov, Yakir} and D. Rohrlich\index{Rohrlich, Daniel}, {\it Quantum Paradoxes:  Quantum Theory for the Perplexed} (Weinheim: Wiley-VCH), 2005, Chap. 10.

\bibitem{GHZ} D. M. Greenberger\index{Greenberger, Daniel}, M. Horne\index{Horne, Michael A.}, and A. Zeilinger\index{Zeilinger, Anton}, ``Going beyond Bell's theorem", in {\it Bell's Theorem, Quantum Theory, and Conceptions of the Universe} [Proceedings of the Fall Workshop, Fairfax, Virginia, October 1988], ed. M. Kafatos (Dordrecht: Kluwer Academic Pub.), 1989, pp. 69-72.

\bibitem{jam} J. Grunhaus, S. Popescu\index{Popescu, Sandu} and D. Rohrlich\index{Rohrlich, Daniel}, ``Jamming nonlocal quantum correlations", {\it Phys. Rev.} {\bf A53} (1996) 3781; D. Rohrlich\index{Rohrlich, Daniel}, ``Three attempts at two axioms for quantum mechanics", in ({\it The Frontiers Collection}) {\it Probability in Physics}, eds. Y. Ben-Menahem and M. Hemmo (Berlin: Springer), 2012, pp. 187-200.  The latter paper notes that jamming arises in quantum mechanics, contrary to what the former paper assumes.

\bibitem{ar} Y. Aharonov\index{Aharonov, Yakir} and D. Rohrlich\index{Rohrlich, Daniel}, {\it op. cit.}, Sect. 18.2.

\bibitem{update} Y. Aharonov\index{Aharonov, Yakir} and L. Vaidman\index{Vaidman, Lev}, ``The two-state vector formalism: an updated review", in {\it Time in Quantum Mechanics}, Volume {\bf 1} [Lecture Notes in Physics {\bf 734}], Second Edition, eds. J. G. Muga, R. S. Mayato and {\'I}. Egusquiza (Berlin: Springer), 2008, pp. 399-447. See also Y. Aharonov\index{Aharonov, Yakir} and D. Rohrlich\index{Rohrlich, Daniel}, {\it op. cit.}, Chap. 18.

\bibitem{weak}Y. Aharonov\index{Aharonov, Yakir}, D. Z. Albert\index{Albert, David Z.}, and L. Vaidman\index{Vaidman, Lev}, ``How the result of a measurement of a component of the spin of a spin-${1\over 2}$ particle can turn out to be 100", {\it Phys. Rev. Lett.} {\bf 60}, 1351 (1988); see also Y. Aharonov\index{Aharonov, Yakir} and D. Rohrlich\index{Rohrlich, Daniel}, {\it op. cit.}, Chaps. 16-17.

\bibitem{arg} N. Argaman, ``Bell's theorem and the causal arrow of time", {\it Am. J. Phys.} {\bf 78}, 1007 (2010).

\bibitem{jb2} J. Bernstein, {\it op. cit.}, pp. 66-67.

\end{references}
\end{document}